# The tunneling Hamiltonian representation of false vacuum decay:
# II. Application to soliton–antisoliton (S-S') pair creation

A. W. Beckwith

Department of Physics and Texas Center for Superconductivity and Advanced Materials at the University of Houston
Houston, Texas 77204-5005, USA

**ABSTRACT**

The tunneling Hamiltonian has proven to be a useful method to treat particle tunneling between different states represented as wavefunctions in many-body physics. Previously, we applied it to a driven sine-Gordon system. Here we apply the method to a generalization of the tunneling Hamiltonian to charge density wave transport problems, in which we consider tunneling between states that are wavefunctionals of a scalar quantum field $\phi$. We derive *I-E* curves that match Zenier curves used to fit data experimentally with wavefunctionals congruent with the false vacuum hypothesis.

# I. INTRODUCTION

The quantum decay of the false vacuum hypothesis[1] has been of broad scientific interest for over two decades. It permits us to *invert* the potential and to treat what was previously a quasi-potential well problem as a potential barrier tunneling between different 'potential' states. The decay of the false vacuum is a potent paradigm for a decay of a *metastable* state to one of *lower* potential equilibrium. We use the generalized Euclidian action procedure previously outlined[2] for a charge density wave (CDW) transport problem; this allows us, for the first time, to obtain a current density expression that matches experimental data sets, as we did in our CDW analysis with soliton-antisoliton (S-S') pairs .

The tunneling Hamiltonian[3,4] involves matrix elements in the transfer of particles between initial and final wave functions. The utility of the functional tunneling Hamiltonian becomes especially apparent since it permits putting potential energy information in the wave functionals and analyzing the kinetics of the evolution between initial and final wavefunctional states. Moreover, a number of experiments on charge density waves and other condensed matter systems suggest quantum decay of the false vacuum, accompanied by the nucleation of soliton domain walls, even when the total action is large. Also the techniques we derive here fits within a wide literature of more abstractly presented treatments of this idea.[5] We also claim that the fixed distance $L$ we obtain between the S-S' components is a de facto quantization condition.[2]

## II USING THE BOGOMIL'NYI INEQUALITY TO MAKE LINKAGE WITH THE FATE OF FALSE VACUUM HYPOTHESIS IN CDW TRANSPORT

We will initiate our inquiry by addressing how the tunneling Hamiltonian ties in with the fate of the false vacuum hypothesis. To do this, we begin with the CDW basics we previously emloyed.

Following J.H. Miller, we use the extended Schwinger model[9] with

$$H = \int_x \left[ \frac{1}{2 \cdot D} \cdot \Pi_x^2 + \frac{1}{2} \cdot (\partial_x \phi_x)^2 + \frac{1}{2} \cdot \mu_E^2 \cdot (\phi_x - \varphi)^2 + \frac{1}{2} \cdot D \cdot \omega_P^2 \cdot (1 - \cos\phi) \right] \quad (1)$$

This lead to us using the thin-wall approximation[10,11] in phase of the form

$$\phi_0 \equiv \pi \cdot [\tanh b(x - x_a) + \tanh b(x_b - x)] \quad 2)$$

**[put Figure 1 about here]**

Let us begin with what the Bogomol'nyi inequality [6] tells us about functionals used in our CDW transport problem. It gives us $L^{-1}$ and fits with the fate of the false vacuum hypothesis which gives us a distinctive $\Delta E$ value.[2]

**[put Figure 2 about here]**

The extended sine Gordon model[11] permits us to write an Euclidian action potential of the form

$$V_E\big|_{\phi \equiv \phi_0} = D \cdot \omega_P^2 \cdot (1 - \cos\phi_0) + \mu_E \cdot (\phi_0 - \Theta)^2 \quad (3)$$

with $\phi_0$ varying in a way for which

$$D \cdot \omega_P^2 >> \mu_E \quad (4)$$

This allowed us to obtain a suitable set of values of $\phi_F$ and $\phi_T$ values of phase, for which

$$\left\{ \frac{\partial \cdot V_E}{\partial \cdot \phi} \bigg|_{\phi \equiv \phi_{T,F}} \right\} = D \cdot \omega_P^2 \cdot \sin \phi_{T,F} + 2 \cdot \mu_E \cdot (\phi_{T,F} - \Theta) \tag{5}$$

is for all purposes zero which gives suitable values of

$$\frac{\partial V}{\partial \phi} = D\omega_p^2 \sin\phi + 2\mu_E(\phi - \theta) \approx D\omega_p^2 \phi + 2\mu_E(\phi - \theta) = 0,$$

$$\Rightarrow \phi_F \approx \left[ \frac{2\mu_E}{D\omega_p^2 + 2\mu_E} \right] \theta \approx \left[ \frac{\varepsilon^+}{\varepsilon^+ + 1} \right] \cdot \theta \approx \varepsilon^+ \tag{6}$$

This is then tied in with the Bogomol'nyi inequality[6] formulation of

$$L_E \approx \left[ L_E |_{\phi = \phi_C} + \frac{4}{4^2} \cdot \frac{D \cdot \omega_P^2}{3!} \cdot (\phi_0^2 - \phi_C^2)^2 \right] + \frac{1}{2} \cdot (\phi_0 - \phi_C)^2 \cdot \{\ \} \tag{7}$$

Due to a *topological* current argument due to $|Q| \to 0$

and

$$\{\ \} \equiv \{\ \}_A - \{\ \}_B \equiv 2 \cdot \Delta E_{gap} \tag{8}$$

where

$$\{\ \}_A \cong D \cdot \omega_P^2 \cdot \cos \phi_F + 2\mu_E \approx D \cdot \omega_P^2 + 2\mu_E \tag{9}$$

and

$$\{\ \}_B \cong \left( \frac{2}{3!} \phi_F \cdot \phi_T \right) D \cdot \omega_P^2 \tag{10}$$

We get a connection with the fate of a false vacuum paradigm [1] and the Bogomil'nyi inequality[2,6] if

$$\frac{(\{\ \})}{2} \equiv \Delta E_{gap} \equiv V_E(\phi_F) - V_E(\phi_T) \tag{11a}$$

$$.009782 \cdot D \cdot \omega_P^2 \equiv \mu_E \Leftrightarrow \phi_F \equiv .11085, \phi_T \equiv 2 \cdot \pi + .00001674$$
$$\Rightarrow \Delta E_{gap} \cong .373 \cdot D \cdot \omega_P^2 \tag{11b}$$

This is (setting $D \cdot \omega_P^2 \equiv 1$ for scaling purposes) akin to what we have when we look at the right hand side of Fig. 1 as well as Fig. 2. We should note that our problem falls apart if we do not satisfy Eq. 11a above. Now, we may specify Eqs. 6 and 7 above as being linked to CDW transport if

$$\Rightarrow \Delta E_{gap} \cong .373 \cdot D \cdot \omega_P^2 \approx L^{-1} \cong \alpha_1 \equiv \alpha_2 \tag{12}$$

and

$$\phi_T \geq \phi_0 \tag{13}$$

$$\Psi_{final} = c_2 \cdot \exp\left(-\alpha_2 \cdot \int d\tilde{x}[\phi_0 - \langle\phi\rangle_1]^2\right) \tag{14}$$

where $\phi_F \equiv <\phi>_1 \cong$ very small value, and $|\phi_T| \geq |\phi_0| \cong 2 \cdot \pi$

where we are assuming *compact support* for the integrand when $\tilde{x} \in \left[-\frac{L}{2}, \frac{L}{2}\right]$

as well as

$$\Psi_{initial} \cong c_1 \cdot \exp\left(-\alpha_1 \cdot \int dx[\phi_0 - \phi_{2\pi}]^2\right) \tag{15}$$

We have $\alpha_2 \cong \alpha_1$ as a convenience in our subsequent calculations in momentum space.

# III ANALYZING THESE WAVE FUNCTIONALS IN MOMENTUM SPACE FOR CDW

We shall now convert into momentum space the action integrals we write as

$$\left(\int L_1 d\tau\right) \to \int dx [\phi_0 - \phi_C]_1^2 \bigg|_{\phi_C \equiv \phi_{C1}} \tag{16}$$

and

$$\left(\int L_2 d\tau\right) \to \int dx [\phi_0 - \phi_C]_2^2 \bigg|_{\phi_C \equiv \phi_{C2}} \tag{17}$$

In the case of CDW this will be when $\phi_{C1}(x)|_1$ *is a nearly 'flat' state indicating pre-nucleation* values of the S-S' pair which we would call a non-nucleated state approaching $\phi_F$ in the situation defined by Figs. 1 and 2, whereas $\phi_{C2}(x)|_2$ is, with regards to a *nearly* fully formed S-S' pair, approaching the $\phi_T(x)$ value as seen in Fig. 2 — with $\phi_T(x)$ being represented by the S-S' pair of height $2 \cdot \pi + \varepsilon^+$ and of width *L*, *where L* is the distance between a S-S'. We assume that $\phi_0 \to \phi_T - \varepsilon^+$ in value and is nearly at that value $\phi_{C2}$. Usually, when we do this, we have that the scaled height $n_1 \cdot 2 \cdot \pi \leq 2 \cdot \pi$ of a S-S' pair with $n_1 \leq 1$ and usually a bit less than 1 in value for $\phi_0 \to \phi_T - \varepsilon^+$, we should write a basis state for S-S' pairs as:[10]

$$\phi(x) = \frac{(2 \cdot \pi)}{L} \cdot \sum_n |\phi(k_n)| \tag{18}$$

which will lead to having a DFT representation of equation 16[10] as

$$\alpha \cdot \int dx [\phi_0 - \phi_C]_{\phi_C \equiv \phi_T}^2 \equiv \left(\frac{2 \cdot \pi}{L}\right)^2 \cdot \sum_n |\phi(k_n)|^2 \tag{19}$$

and a DFT representation of the equation 17 [10] as

$$\alpha \cdot \int dx [\phi_0 - \phi_C]^2_{\phi_C \equiv \phi_F} \equiv \left(\frac{2 \cdot \pi}{L}\right)^2 \cdot \sum_n (1 - n_1^2) \cdot |\phi(k_n)|^2 \tag{20}$$

when $\phi_C \equiv \phi_T$ with a S-S' sub box height $n_1 \cdot (2 \cdot \pi)$ being contained within and evolving to the final configuration box S-S' box of length $L$ and height about the value of $(2 \cdot \pi)$. Thus, we may write

$$\Psi_{final} \equiv C_2 \cdot \exp\left(-\left(\frac{\alpha}{L}\right) \cdot (2 \cdot \pi)^2 \cdot \sum_n |\phi(k_n)|^2\right) \tag{21}$$

In addition, we have

$$\Psi_{initial} = C_1 \cdot \left(-\left(\frac{\alpha}{L}\right) \cdot (2 \cdot \pi)^2 \cdot \sum_n (1 - (n1)^2) \cdot |\phi(k_n)|^2\right) \tag{22}$$

as well as a momentum representation of path integrals via

$$\wp \cdot \phi(x) \equiv \prod_n \frac{1}{L} \cdot \left\{\frac{d}{d \cdot k_n} \cdot [\exp(-i \cdot k_n \cdot x) \cdot \phi(k_n)]\right\} \cdot d \cdot k_n \tag{23}$$

as well as

$$\frac{\delta \cdot \psi_{1,2}}{\delta \cdot \phi(x)} \equiv \sum_n \frac{\delta \cdot \psi_{1,2}}{\delta \cdot \phi(k_n)} \cdot \left(\left(\frac{\partial \cdot \phi(k_n)}{\partial \cdot k_n}\right) \cdot \left(\frac{\partial \cdot \phi(x)}{\partial \cdot k_n}\right)^{-1} \equiv f(k_n)\right) \tag{24}$$

and, assuming $\alpha_1 \cong \alpha_2 \equiv \alpha \approx \frac{1}{L}$, as well as assuming that the geometry of Fig. 2 holds[2]

$$\Psi_{initial} \equiv \Psi_1(\phi_F) \cong c_1 \cdot \exp\left(-\alpha \cdot \int d\tilde{x} [\phi_F]^2\right) \tag{26}$$

as well as

$$\Psi_{final} \equiv \Psi_2(\phi_T) \cong c_2 \cdot \exp\left(-\alpha \cdot \int d\tilde{x} [\phi_T]^2\right) \tag{27}$$

where we can state the exponential terms of the initial and final wavefunctionals to have

$$\phi(k_{n_{a1},a2}) \equiv \phi(k) \equiv \phi(k_n) = \sqrt{\frac{2}{\pi}} \cdot \frac{\sin(k_n L/2)}{k_n} \tag{29}$$

## IV. ELIMINATION OF CROSS TERMS IN $T_{IF}$

We should note that the fact that we look at only at a fixed value of momentum allows[10]

$$\wp \cdot \phi(x) \equiv \frac{1}{L} \cdot \left\{ \frac{d}{d \cdot k_N} \cdot [\exp(-i \cdot k_N \cdot x) \cdot \phi(k_N)] \right\} \cdot d \cdot k_N \tag{30}$$

so[10]

$$\Psi_1 \to C_1 \cdot \exp\left(-\left(\frac{\alpha}{L}\right) \cdot (2 \cdot \pi)^2 \cdot |\phi(k_N)|^2\right) \tag{31}$$

and[10]

$$\Psi_2 \to C_2 \cdot \exp\left(-\left(\frac{\alpha}{L}\right) \cdot (2 \cdot \pi)^2 \cdot (1 - n_1^2) \cdot |\phi(k_N)|^2\right) \tag{32}$$

with the 'normalization' so that for $i = 1,2$ we may write

$$C_i = \frac{1}{\sqrt{\sqrt{\frac{L^2}{2 \cdot \pi}} \int_0^{\cdot} \exp(-2 \cdot \{\}_i \cdot \phi^2(k_N)) \cdot d\phi(k_N)}} \tag{33}$$

where for the different wavefunctionals we evaluate for $i = 1,2$ via the error function[12]

$$\int_0^{\sqrt{\frac{L^2}{2\cdot\pi}}} \Psi_i^2 \cdot d\phi(k_N) = 1 \tag{34}$$

$$\int_0^b \exp(-a\cdot x^2)\,dx = \frac{1}{2}\sqrt{\frac{\pi}{a}}\cdot \text{erf}(b\cdot\sqrt{a}) \tag{35}$$

So

$$\frac{\delta^2 \cdot \psi_{1,2}}{\delta\phi(x)^2} \equiv -2\cdot C_{1,2}\cdot \sum_{na_1,na_2} \frac{\delta}{\delta\cdot\phi(k_{na_2})}\{\ \}_{1,2}\cdot f(k_{na_1})\cdot f(k_{na_2}) \tag{36}$$

where we can set

$$\{\ \}_{1,2} \equiv [\ ]_{1,2}\phi(k_{na_1})\cdot \exp(-[\ ]_{1,2}\cdot \phi^2(k_{na_1})) \tag{37}$$

so we then will be looking at

$$\frac{\delta^2 \cdot \psi_{1,2}}{\delta\phi(x)^2} \equiv 2\cdot C_{1,2}\cdot \sum_{na_1} \frac{\delta}{\delta\cdot\phi(k_{na_1})}\ [\{\ \}]_{1,2}\cdot f^2(k_{na_1}) \tag{38}$$

$$[\{\ \}]_{1,2} \equiv [\ ]_{1,2}^2\phi^2(k_{na_1})\cdot \exp(-[\ ]_{1,2}\cdot \phi^2(k_{na_1})) \tag{39}$$

as well as looking at converting the integration with respect to phase $\phi(x)$ to $dk_N$ (with momentum as $k_N$) with the other terms not contributing with

$$\frac{\delta^2 \cdot \psi_{1,2}}{\delta\phi(x)^2} \equiv 2\cdot C_{1,2}\cdot [\ ]_{1,2}^2 \phi^2(k_{N_1})\cdot \exp(-[\ ]_{1,2}\cdot \phi^2(k_N))\cdot f^2(k_N) \tag{40}$$

and this is mainly due to non-zero pole singularities appearing in the momentum space

$$f(k_N) \equiv \exp(i\cdot k_N\cdot x)\cdot \left[\frac{\frac{k_N L}{2}\cdot \cos\left(\frac{k_N L}{2}\right) - \sin\left(\frac{k_N L}{2}\right)}{\cos\left(\frac{k_N L}{2}\right) - \left[i\cdot k_N\cdot x + \frac{1}{k_N\cdot L}\right]\cdot \sin\left(\frac{k_N L}{2}\right)}\right] \tag{41}$$

with all but the n represented as N contribution in the wavefunctionals ignored so we can then look at an integral of the form for $T_{IF}$ as having an absolute magnitude of

$$|T_{IF}| \approx \cdot \frac{2}{2 \cdot m^*}\left(n_1^2 - \frac{n_1^4}{2}\right) \cdot C_1 \cdot C_2 \cdot \left(\cosh\left(2\sqrt{\frac{x}{2L}} - \sqrt{\frac{L}{2x}}\right)\right) \cdot e^{-\alpha \cdot L\left[n_1^2 \cdot \frac{L}{2x}\right]} \quad (42)$$

where we are assuming that we are using a scaling of $\hbar \equiv 1$, and which if we use $n_1 \cong 1 - \varepsilon^+$ becomes[10]

$$|T_{IF}| \approx \frac{C_1 \cdot C_2}{m^*} \cdot \left(\cosh\left(2\sqrt{\frac{x}{2L}} - \sqrt{\frac{L}{2x}}\right)\right) \cdot e^{-\alpha \cdot L\left[\frac{L}{2x}\right]} \quad (43)$$

a complex valued integration which would vanish if the imaginary contribution of $T_{IF}$ were ignored. So then we are working with a current which is the magnitude of a residue calculation[10] where we have

$$T_{IF} \approx \int \frac{f(k_N)}{g(k_N - i \cdot (value))} \cdot dk_N \quad (44)$$

where the numerator $f$ and denominator $g$ are analytic complex valued function. We should note that this $T_{IF}$ would be zero if we were not counting imaginary root contributions to the functional integral for our tunneling Hamiltonian. Note, that the S-S' pairs will form a current, and this will occur when we have condensed electrons tunneling through a pinning gap at the Fermi surface in order to accelerate the CDW with respect to an electric field. Fig. 3 captures the essence of this current behavior[12] mainly because we have only modeled a

**[put Figure 3 about here]**

non zero current composed of S-S' pairs when $E_{DC} \geq E_T$. Note that the Bloch bands are tilted by an applied electric field when we have $E_{DC} \geq E_T$ leading to a S-S' pair as shown in Fig. 2[13]. The slope of the tilted band structure is given

**[put Figure 4 about here ]**

by $e^* \cdot E$ and the separation between the S-S' pair is given by:

$$L = \left(\frac{2 \cdot \Delta_s}{e^*}\right) \cdot \frac{1}{E} \tag{45}$$

So, that, then, we have $L \propto E^{-1}$. If we consider a Zener diagram of CDW electrons with tunneling only happening when $e^* \cdot E \cdot L > \varepsilon_G$ where $e^*$ is the effective charge of each condensed electron and $\varepsilon_G$ being a pinning gap energy, we have that Fig. 3 permits us to write[10]

$$\frac{L}{x} \equiv \frac{L}{\bar{x}} \cong c_v \cdot \frac{E_T}{E} \tag{46}$$

Here, $c_v$ is a proportionality factor included to accommodate the physics we obtain via a given spatial (for a CDW 'chain') harmonic approximation of

$$\bar{x} = \bar{x}_0 \cdot \cos(\omega \cdot t) \Leftrightarrow m_{e^-} \cdot a = -m_{e^-} \cdot \omega^2 \cdot \bar{x} = e^- \cdot E \Leftrightarrow \bar{x} = \frac{e^- \cdot E}{m_{e^-} \omega^2}$$

Realistically, we have that $L \gg \bar{x}$, where we assume that $\bar{x}$ is an assumed reference point an observer picks to measure where a S-S' pair is on an assumed one-dimensional chain of impurity sites. All of this allows us to write the given magnitude of $|T_{IF}|$ as directly proportional to a current formed of S-S' pairs, which is further approximated to be[10]

$$I \propto \tilde{C}_1 \cdot \left[ \cosh\left[ \sqrt{\frac{2 \cdot E}{E_T \cdot c_V}} - \sqrt{\frac{E_T \cdot c_V}{E}} \right] \right] \cdot \exp\left( -\frac{E_T \cdot c_V}{E} \right) \qquad (47)$$

where we are using the normalization constants of the wave functionals via

$$\tilde{C}_1 \equiv \frac{C_1 \cdot C_2}{m^*} \qquad (48)$$

which is a great refinement upon the phenomenological Zenier current[7] expression

$$I \propto G_P \cdot (E - E_T) \cdot \exp\left( -\frac{E_T}{E} \right) \; if \, E > E_T \qquad (49)$$

**[Put Fig. 5 about here]**

Otherwise, we are restricting ourselves to ultra fast transitions of CDW which is realistic and in sync with how our wave functionals used are formed in part by the fate of the false vacuum hypothesis.

## V. COMPARISON WITH GENERALIZATION OF SWINGERS RESULT

We shall now refer to a 1999 paper by Qiong-gui Lin,[8] who came up with a general rule with respect to the probability of electron-positron pair creation in D+1 dimensions, with D varying from one to three, leading to in the case of a pure electric field:

$$w_E = (1 + \delta_{d3}) \cdot \frac{|e \cdot E|^{(D+1)/2}}{(2 \cdot \pi)^D} \cdot \sum_{n=1}^{\infty} \frac{1}{n^{(D+1)/2}} \cdot \exp\left( -\frac{n \cdot \pi \cdot m^2}{|e \cdot E|} \right) \qquad (50)$$

If D is set equal to three, we get (after setting $e^2, m \equiv 1$)

$$wIII(E) = \frac{|E|^2}{(4 \cdot \pi^3)} \cdot \sum_{n=1}^{\infty} \frac{1}{n^2} \cdot \exp\left(-\frac{n \cdot \pi}{|E|}\right) \tag{51}$$

which, if graphed gives a comparatively flattened curve compared w.r.t. to what we get if D is set equal to one ( after setting $e^2, m \equiv 1$ )

$$wI(E) = \frac{|E|^1}{(2 \cdot \pi^1)} \cdot \sum_{n=1}^{\infty} \frac{1}{n^1} \cdot \exp\left(-\frac{n \cdot \pi}{|E|}\right) \equiv -\frac{|E|}{2 \cdot \pi} \cdot \ln\left[1 - \exp\left(-\frac{\pi}{E}\right)\right] \tag{52}$$

which is far more linear in behavior for an e field varying from zero to a small numerical value. We see these two graphs in Fig. 6,

**[Put Figure 6 about here]**

and we note that this is indicating that as dimensionality drops, we have a steady progression toward linearity. The three dimensional result as given by Lin is merely the Swinger result[16] given in the 1950s. When we have $D = 1$, we are approaching behavior very similar to what we obtain with the analysis completed for the S-S' current argument just presented, with the main difference lying in a threshold electric field that is cleanly represented by our graphical analysis, which is a major improvement in the prior curve fitting exercised used in 1985 to curve-fit data.[7]

## VI. CONCLUSION

We have managed to link the fate of the false vacuum hypothesis[1] with a wave functional formalism,[2] which permits gaussian approximations of potential energy contributions[2] to the extended swinger model[11] in CDW dynamics. In addition, we have, for the first time, used this method to construct an **I-E** curve that improves upon a prior Zener curve-fitting approximation used in 1985[7] to obtain a close fit with experimental data sets. This is important since it establishes that we need a pinning gap analysis[2,13] with S-S' pairs to make sense of what was previously a result that did not have a rigorous derivation.[7] In addition, we also have shown that this procedure fits well within an Euclidian least action argument pioneered by Sidney Coleman[1] via use of the vanishing of a topological charge[2] for a S-S' pair traversing a pinning gap.[13] This establishes, via use of the Bogomil'myi inequality[2,6] that we can think of S-S' pair transport as having almost instantaneous jumps[10] (seen experimentally all the time) as well has having a well-specified width,[2] which can be viewed as part of a quantization condition for this problem.[2] Finally, we have shown how the **I-E** curve we derived has similarities with the behavior of nucleation of an electron-positron pair to the minimum dimensionality, as predicted by Swinger[14] when we reduce the dimensionality of the analyzed results Lin[8] gave us, which adds credence to our quasi one-dimensional analysis of CDW dynamics.[2,10]

# FIGURE CAPTIONS

**Fig 1:** Evolution from an initial state $\Psi_i[\phi]$ to a final state $\Psi_f[\phi]$ for a double-well potential (inset) in a 1-D model, showing a kink-antikink pair bounding the nucleated bubble of true vacuum. The shading illustrates quantum fluctuations about the classically optimum configurations of the field $\phi_i = 0$ and $\phi_f(x)$, while $\phi_0(x)$ represents an intermediate field configuration inside the tunnel barrier.

**Fig 2.** Fate of the false vacuum representation of what happens in CDW. This shows how we have a difference in energy between false and true vacuum values and how this ties in with our Bogomil'nyi inequality.

**FIG 3.** The above figures represents the formation of soliton-anti soliton (S-S') pairs along a chain. The evolution of phase is spatially given by

$$\phi(x) = \pi \,[\tanh b(x-x_a) + \tanh b(x_b - x)].$$

**FIG 4.** This is a representation of 'Zener' tunneling through pinning gap with band structure tilted by applied E field.

**FIG 5.** Experimental and theoretical predictions of current values. The dots represent a Zenier curve fitting polynomial, whereas the blue circles are for the S-S' transport expression derived with a field theoretic version of a tunneling Hamiltonian.

**FIG 6.** Two curves representing probabilities of the nucleation of an electron-positron pair in a vacuum. $wI(E)$ is a nearly-linear curve representing a 1+1 dimensional system, whereas the second curve is for a 3 + 1 dimensional physical system and is far less linear in behavior.

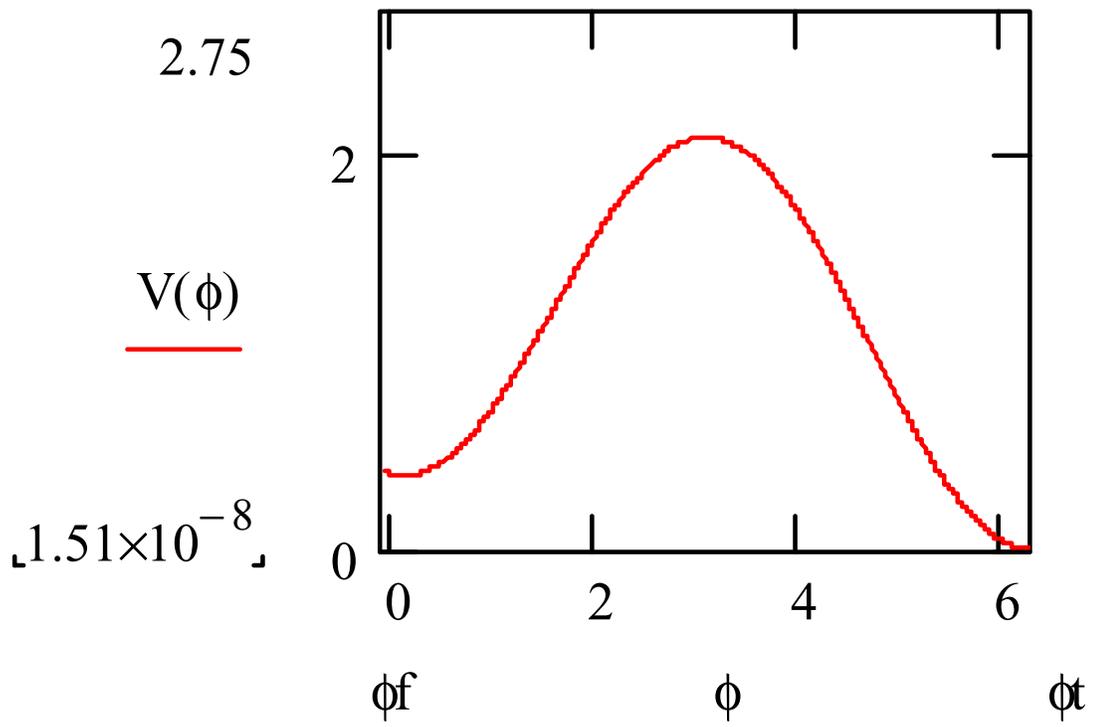

FIGURE 1

BECKWITH

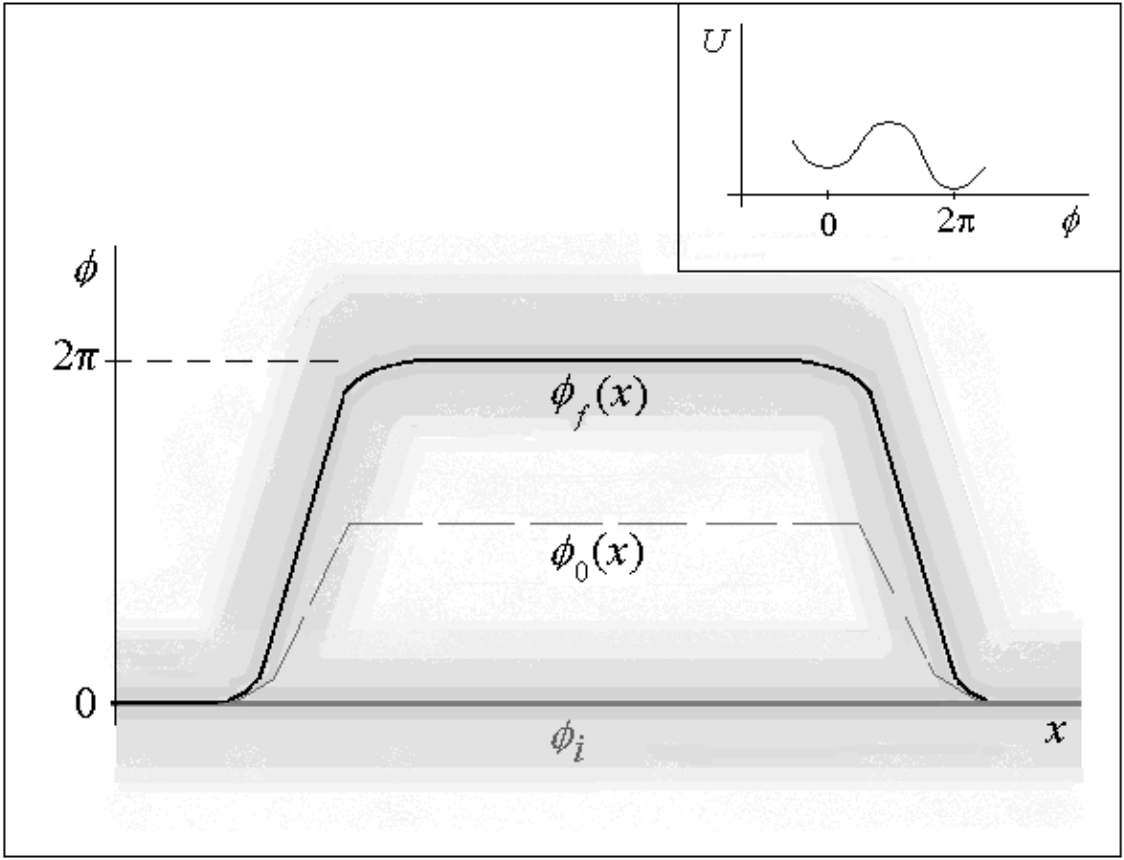

FIGURE 2

BECKWITH

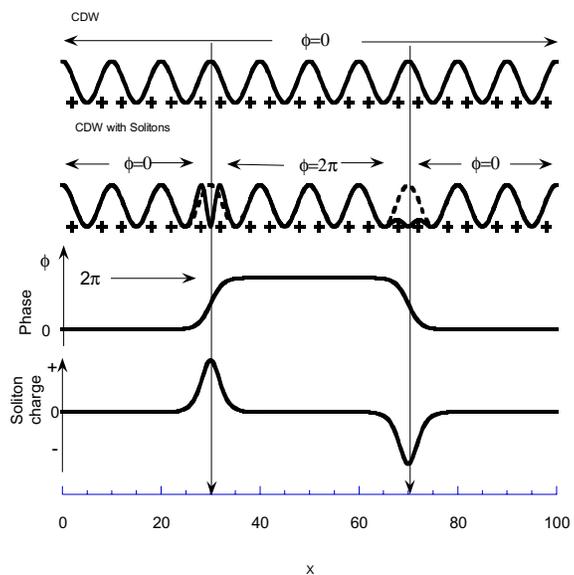

FIGURE 3

BECKWITH ET AL

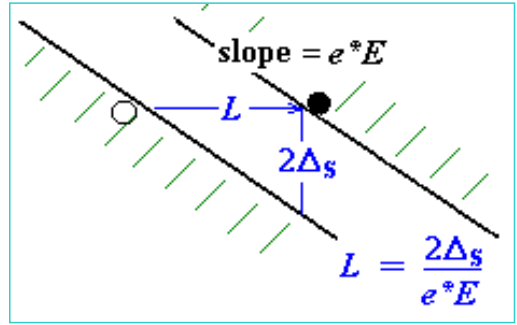

FIGURE 4

BECKWITH

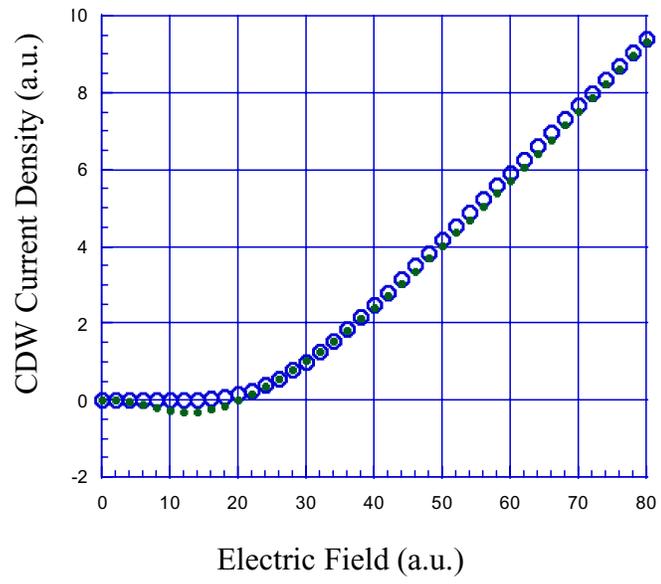

FIGURE 5

BECKWITH

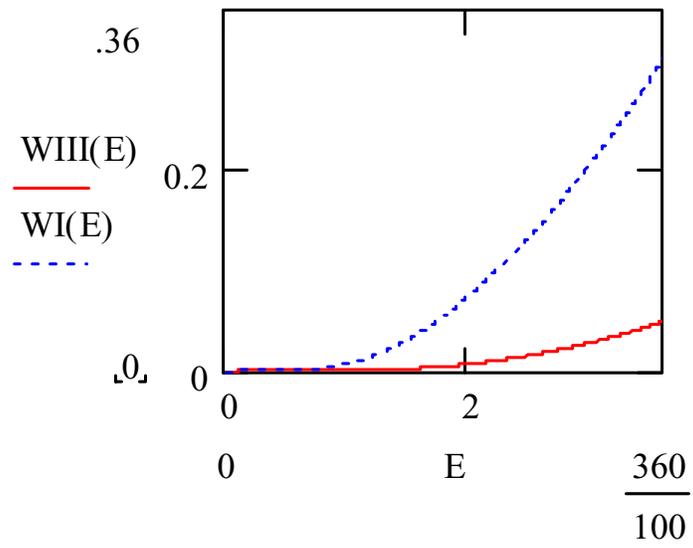

FIGURE 6

BECKWITH